\documentclass{iopart}
\usepackage{graphicx}
\usepackage{color}
\usepackage{amssymb}

\begin{document}

\newcommand{\be}{\begin{equation}}
\newcommand{\ee}{\end{equation}}
\renewcommand{\S}{\mathfrak{S}}
\newcommand{\C}{\mathcal{C}}

\title[Combinatorial problems in the semiclassical approach to transport]{Combinatorial problems in the semiclassical approach to quantum chaotic transport}

\author{Marcel Novaes}

\address{Departamento de F\'isica, Universidade Federal de S\~ao Carlos, S\~ao Carlos,
SP, 13565-905, Brazil}


\begin{abstract}
A semiclassical approach to the calculation of transport moments $M_m={\rm Tr}[(t^\dag
t)^m]$, where $t$ is the transmission matrix, was developed in \cite{epl} for chaotic
cavities with two leads and broken time-reversal symmetry. The result is an expression
for $M_m$ as a perturbation series in $1/N$, where $N$ is the total number of open
channels, which is in agreement with random matrix theory predictions. The coefficients
in this series were related to two open combinatorial problems. Here we expand on this
work, including the solution to one of the combinatorial problems. As a by-product, we
also present a conjecture relating two kinds of factorizations of permutations.

\end{abstract}


\section{Introduction}

Wave scattering in systems whose classical dynamics is chaotic displays universal
statistics, as first observed in experiments with coherent electronic transport in
ballistic quantum dots \cite{fluc1,fluc2}. We consider a chaotic cavity attached to two
ideal leads having $N_1$ and $N_2$ open channels. This is described by a $N\times N$
scattering matrix $S$, where $N=N_1+N_2$ is the total number of channels. The $S$ matrix
is always unitary, reflecting conservation of charge (if time-reversal symmetry is
present, it is also symmetric). Following similar considerations in the field of quantum
chaos, it was soon realized that these statistical properties are described by the theory
of random matrices (RMT) \cite{uzy,uzy3,jala1,jala2,rmt}. This theory neglects all
particularities a system may have and replaces $S$ by a random unitary matrix
\cite{others} (unitary symmetric if time-reversal symmetry is present).

Let $t$ be the $N_1\times N_2$ transmission block of $S$. The transport moments $M_m={\rm
Tr}[(tt^\dag)^m]$ carry information about the scattering process (so-called linear
statistics). The $m=1$ moment is called the conductance, and the $m=2$ moment is related
to the shot-noise. Within RMT, the average value of $M_m$ has long been considered.
Initially, only perturbative results were obtained \cite{pert,pert2}, valid to leading or
next-to-leading order in $1/N$, and only for the first moments. In recent years the
connection with the Selberg integral was properly realized \cite{savin2005} and this
eventually allowed the calculation of all $M_m$ to be carried out for arbitrary values of
$N_1$ and $N_2$, for both symmetry classes \cite{prb78mn2008,savin2009} (see also
\cite{Vivo}).

It has always been a central problem to derive these universal statistics from a
semiclassical approximation. Quantum universality must emerge semiclassically as a result
of action correlations: there must exist sets of scattering trajectories having nearly
the same total action, so that they may interfere constructively. Conductance and
shot-noise were considered at leading order \cite{espalha,geisel} and later to all orders
\cite{prl96sh2006,shot}. Higher moments were treated to leading order in
\cite{jpa41gb2008} and to next-to-leading order in \cite{njp13gb2011}.

Recently, works by this author \cite{epl} and by Berkolaiko and Kuipers \cite{also} have
presented semiclassical calculations of all $M_m(N_1,N_2)$ valid to all orders in $1/N$.
The approach developed in \cite{epl}, for systems with broken time-reversal symmetry,
related the transport problem to correlated periodic orbits and then arrived at two
combinatorial problems that were left open. We present here the solution to one of these
problems. The other one, which remains open, is to determine the number of solutions to
the factorization $(12\cdots E)=QP$ of the cyclic permutation under some conditions on
the factors. Interestingly, the approach developed in \cite{also} (which firmly
establishes the equivalence of semiclassics and RMT for all symmetry classes) requires
the solution of another factorization problem, involving so-called primitive
factorizations \cite{merola}. Primitive factorizations are important within the symmetric
group and the unitary group \cite{novak}. In the present work, we compare our
semiclassical result to RMT and to the result of \cite{also}, arriving at a conjecture
relating both kinds of factorization problems.

The paper is organized as follows. In Section 2 we present some combinatorial
preliminaries and advance our conjecture. In Section 3 we introduce the semiclassical
theory of quantum chaotic transport. In Section 4 we review the work summarized in
\cite{epl}, and solve one of the combinatorial problems left open in that paper. In
Section 5 we compare our theory to RMT and to the different semiclassical approach of
\cite{also}. We present some conclusions in Section 6.

\section{Combinatorial preliminaries and a conjecture}

We shall make use of several combinatorial results, so we start with a brief revision of
the ones we need. We are rather concise since these facts are widely known. We also
present the conjecture that arises from our work.

A sequence of positive integers $\lambda=(\lambda_1,\lambda_2,\ldots\lambda_\ell)$ with
$\lambda_i\geq \lambda_{i+1}$ is called a partition of $m$ if $\sum_i\lambda_i=m$. This
is denoted by $\lambda\vdash m$. Each of these integers is called a part, and the total
number of parts is called the length of the partition, $\ell(\lambda)$. It is also usual
to write a partition in a frequency representation, i.e. as
$\lambda=1^{a_1}2^{a_2}\cdots$ where $a_j$ is the number of times the part $j$ appears in
$\lambda$. For example, $(3,2,2,1)=12^23$ is a partition of $8$.

The quantity \be{m \choose \lambda}=\frac{m!}{\lambda_1!\lambda_2!\cdots
a_1!a_2!\cdots}\ee counts in how many ways we can partition the set $\{1,\ldots,m\}$ into
subsets having the parts of $\lambda$ as cardinalities. The number of set partitions into
exactly $k$ non-empty subsets is the Stirling number of the second kind, \be
S(m,k)=\sum_{\ell(\lambda)=k}{m \choose \lambda}.\ee These numbers satisfy \be
\sum_{k=0}^m S(m,k)[x]_k=x^m,\ee where \be [x]_k=x(x-1)\cdots(x-k+1)\ee is the falling
factorial.

Let $\S_m$ be the group of all $m!$ permutations of $m$ symbols. When a given permutation
acts, it divides the $m$ elements into orbits called cycles. For example, the permutation
$(123)(45)(6)$ acts on the set of the first $6$ integers and produces $3$ cycles: $1\to
2\to 3\to 1$ is a cycle of length three, $4\to 5\to 4$ is a cycle of length two and $6$
is a fixed point. Fixed points are often omitted when writing a permutation, so
$(123)(45)(6)\equiv (123)(45)$. By convention, all cycles start with their smallest
element and they are written in increasing order of the first element.

The lengths of the cycles of a permutation $\pi\in\S_m$ form a partition of $m$ called
the cycle type of $\pi$. The cycle type of $(123)(45)(67)$ is $2^23$. A permutation whose
cycle type is $1^{m-2}2$, which simply exchanges two elements, is called a transposition.

Cycle type is preserved under the action of conjugation, i.e. $\pi$ and
$\sigma\pi\sigma^{-1}$ have the same cycle type for any $\sigma$. Therefore, the group
$\S_m$ may be divided into conjugacy classes $\C_\lambda$, each class being determined by
a partition of $m$ which is the cycle type of its elements. The number of elements in
$\C_\lambda$ is \be |\C_\lambda|={m \choose \lambda}\prod_i(\lambda_i-1)!.\ee The number
of elements of $\S_m$ which have exactly $k$ cycles, of whatever type, is the (unsigned)
Stirling number of the first kind, \be s(m,k)=\sum_{\ell(\lambda)=k}|\C_\lambda|.\ee
These numbers satisfy \be\label{ris} \sum_{k=0}^m s(m,k)x^k=[x]^m,\ee where \be
[x]^k=x(x+1)\cdots(x+k-1)\ee is the rising factorial.

Enumerating factorizations of permutations is an important problem, with connections to
many other areas. A classical factorization problem is the following. Determine the
number of permutation pairs $(\sigma,\tau)$, with $\sigma\in\C_\alpha$ and
$\tau\in\C_\beta$, such that $\sigma\tau=\pi$ for a fixed permutation $\pi\in\C_\lambda$.
The number of solutions to this problem, denoted $C^\lambda_{\alpha,\beta}$, is called
the connection coefficient of the permutation group. We define \be
C^\lambda_{a,b}=\sum_{\ell(\alpha)=a}\sum_{\ell(\beta)=b}C^\lambda_{\alpha,\beta}\ee as
the solution to the factorization problem where only the numbers of cycles of the factors
are specified. For example, the only element of $\C_{1^m}$ is the identity and its
factorizations are all of the kind $1=\sigma \sigma^{-1}$; thus, it is clear that
$C^{1^m}_{a,b}=s(m,a)\delta_{a,b}.$

A semiclassical calculation of transport moments was developed by Berkolaiko and Kuipers
in \cite{also}, which is different from the one discussed here and in \cite{epl}. They
encounter the following factorization problem: when $\pi=(s_1t_1)\cdots(s_dt_d)$ is a
product of $d$ transpositions with $t_j>s_j$ and $t_k\geq t_j$ for all $k>j$, this is
called a {\em primitive} factorization of $\pi$ of depth $d$. Primitive factorizations
were introduced in \cite{merola} and have been further discussed in \cite{novak}. The
number of such factorizations for a given permutation $\pi$ is denoted by $p_d(\pi)$. It
depends only on the cycle type of $\pi$, so it can be denoted equivalently by
$p_d(\omega)$ if $\pi\in\C_\omega$. This quantity can be obtained by means of a
recurrence relation and this allows the equivalence between RMT and semiclassics to be
firmly established for all symmetry classes \cite{also}.

In this paper we define a new kind of factorization problem, which plays a central role
in our semiclassical approach to transport moments. In order to state it, we need some
notation. Given a permutation $P$, call $P_1$ its first cycle, i.e. the one that contains
`$1$', and let $\{P_1\}$ be the set of elements of $P_1$. A cycle whose elements are in
increasing order is called an increasing cycle. Given a set $s$, let the {\em
restriction} of a permutation $Q$ to $s$, denoted $\left.Q\right|_s$, be the permutation
obtained by simply erasing from the cycle representation of $Q$ all symbols not in $s$.
For example, $\left.(123)(45)\right|_{\{1,3,5\}}=(13)(5)$. The cycle type of such a
permutation is determined by taking the lengths of its cycles as a partition of the
cardinality of $s$.

{\bf Definition:} {\em We denote by $\Xi(m,E,V;\omega)$ the number of solutions in the
permutation group $\S_E$ to the factorization $(12\cdots E)=QP$ of the complete
increasing cycle, which satisfy the following conditions: i) $P$ has $V+1$ cycles and no
fixed points; ii) its first cycle $P_1$ is increasing and of size $m$; iii) all cycles of
$Q$ have at least one element in common with $P_1$; iv) the restriction of $Q$ to
$\{P_1\}$ has cycle type $\omega$}.

For example, the factors $Q=(1432)(5)$ and $P=(135)(24)$ provide a factorization of
$(12345)$ of the kind described above, with $V=2$, $m=3$ and $\omega=(2,1)$. The factors
$Q=(1)(253)(46)$ and $P=(124)(365)$ provide another example at $E=6$, in which $V=2$,
$m=3$ and $\omega=1^3$. Finally, at any value of $E$, the factorization where $Q=1$ and
$P=(12\cdots E)$ is also of this kind, with $V=1$, $m=E$ and $\omega=1^E$. As
counterexamples, $Q=(1)(24)(3)$ and $P=(12)(34)$ are not acceptable as factors at $E=4$
because not all cycles of $Q$ intersect $P_1$, while the factorization with $Q=P=(132)$
is not acceptable because $P_1$ is not increasing.

Comparing our semiclassical expression with the one in \cite{also}, we relate our
factorization problem and the primitive factorization problem through our

{\bf Conjecture:} {\em For every $d\ge 0$ and every $\omega\vdash m$,} \be
\sum_{V=0}^d(-1)^{V+d}\Xi(m,V+m+d,V;\omega) =|\C_{\omega}|p_d(\omega).\ee

It is rather surprising that these two factorization problems, at first sight completely
different, seem to be actually closely related. It would be interesting to have a direct
combinatorial proof of this.

One consequence of this conjecture is that \be\label{conseq} \sum_{\omega\vdash
m}\sum_{V=0}^{d}(-1)^{V+d}\Xi(m,V+m+d,V;\omega)=S(m+d-1,m-1).\ee In order to see this,
notice that \be \sum_{\omega\vdash
m}|\C_\omega|p_d(\omega)=\sum_{\pi\in\S_m}p_d(\pi)=:P(m,d)\ee is simply the cardinality
of the set of all primitive factorizations of depth $d$ in $\S_m$. These factorizations
can be categorized into two types: those which are also primitive factorizations of depth
$d$ in $\S_{m-1}$ and those which are obtained from a primitive factorization of depth
$d-1$ in $\S_{m}$ by appending a factor $(sm)$ at the end, for some $s$. There are $m-1$
possible such factors. Therefore, the cardinality satisfies the recurrence relation \be
P(m,d)=P(m-1,d)+(m-1)P(m,d-1).\ee The Stirling numbers of second kind satisfy \be\fl
S(m+d-1,m-1)=S(m+d-2,m-2)+(m-1)S(m+d-2,m-1),\ee and it is easy to see that
$P(m,1)=S(m,m-1)$ and $P(2,d)=S(d+1,1)$. The result (\ref{conseq}) is thus proved.

\section{Semiclassical Approximation to Transport Moments}

In the semiclassical limit $\hbar\to 0$, $N\to\infty$, the matrix elements of $t$ may be
approximated \cite{c3hub1993a,c3hub1993b} by \be
t_{oi}\approx\frac{1}{\sqrt{T_H}}\sum_{\gamma:i\to o}A_\gamma e^{iS_\gamma/\hbar},\ee
where the sum is over trajectories starting at incoming channel $i$ and ending at
outgoing channel $o$. The phase $S_\gamma$ of trajectory $\gamma$ is its action and the
amplitude $A_\gamma$ is related to its stability. The prefactor contains the Heisenberg
time $T_H$, which equals $N$ times the classical dwell time, i.e. the average time a
particle spends in the cavity (inverse decay rate).

Expanding the trace, transport moments become \be\label{bigsum}
M_m\approx\frac{1}{T_H^m}\prod_{j=1}^m\sum_{i_j,o_j}\sum_{\gamma_j,\sigma_j} A_{\gamma}
A_{\sigma}^* e^{i(S_\gamma-S_\sigma)/\hbar}.\ee The sum involves two sets of $m$
trajectories, the $\gamma$'s and the $\sigma$'s. $A_\gamma=\prod_j A_{\gamma_j}$ is a
collective stability and $S_\gamma=\sum_j S_{\gamma_j}$ is a collective action, and
analogously for $\sigma$. The result of (\ref{bigsum}) is in general a strongly
fluctuating function of the energy, so a local energy average is introduced. When this
averaging is performed in the stationary phase approximation, it selects those sets of
$\sigma$'s that have almost the same collective action as the $\gamma$'s.

Most importantly, the structure of the trace implies that these two sets of trajectories
connect the channels in a different order, and we can arrange it so that $\gamma_j$ goes
from $i_j$ to $o_j$, while $\sigma_j$ goes from $i_{j}$ to $o_{j+1}$. In other words,
when we consider only the labels on the channels, trajectories $\gamma$ implement the
identity permutation, while trajectories $\sigma$ implement $c_m=(12\cdots m)$.

In the past 10 years \cite{martin} it has been established that the way these action
correlations are produced is as follows: each $\sigma$ must follow closely a certain
$\gamma$ for a period of time, and some of them exchange partners at what is called an
encounter. An $r$-encounter is a region where $r$ pieces of trajectories run nearly
parallel and $r$ partners are exchanged. The two sets of trajectories are thus nearly
equal, differing only in the negligible encounter regions. In particular, this implies
$A_{\gamma}A_{\sigma}^*=|A_{\gamma}|^2$. This theory has been presented in detail in
\cite{haakepre,njp9sm2007}. We consider only systems for which the dynamics is not
invariant under time-reversal. Hence, a $\sigma$ trajectory never runs in the opposite
sense with respect to a $\gamma$ trajectory.

Every correlated pair of trajectory sets contributing to (\ref{bigsum}) may be
represented by a diagram, which consists in reducing the encounters to points (vertices)
and drawing the pieces of trajectory connecting them as simple edges (when in reality
they can be extremely convoluted). Doing this erases the information about how exactly
the partner trajectories are exchanged at the encounters. However, it is known that the
contribution of a diagram does not depend on this detailed information, but only on the
total numbers of edges and vertices.

The way this comes about (see \cite{njp9sm2007} for details) is roughly as follows.
First, a rule is used \cite{espalha} that says that a sum over trajectories connecting
any given channels may be replaced by a time integral, \be \sum_{\gamma:i\to
o}|A_\gamma|^2=\int_0^\infty dT e^{-NT/T_H}=\frac{T_H}{N}.\ee Such a factor arises for
each of the edges. On the other hand, integration over all possible action differences at
encounters results that each $r$-encounter produces a factor $-N/T_H^r$. The total power
of $T_H$ becomes $0$ because adding the value of $r$ for all encounters produces $E-m$,
which cancels with the factor from the edges and the denominator in (\ref{bigsum}). In
the end, it can be shown that the total contribution to (\ref{bigsum}) of a diagram with
$V$ vertices and $E$ edges is simply $(-1)^VN^{V-E}$.

It is also necessary to consider how the endpoints of a diagram may be distributed within
the leads. As we shall see, a diagram may require that some channels coincide. If the
number of {\em distinct} incoming and outgoing channels in a diagram are $m_1$ and $m_2$,
respectively, then there are \be [N_1]_{m_1}[N_2]_{m_2}\ee many ways of assigning them
among the possible ones existing in the leads, where $[N]_m$ is the falling factorial.
Finally, if $k$ different $\sigma$'s start and end at the same channels, their identity
may be exchanged without affecting the semiclassical contribution. Therefore, in such a
case the contribution of the diagram must be multiplied by $k!$. Equivalently, we may
define {\em labeled} diagrams as diagrams with a fixed choice of $\sigma$'s, and count
labeled diagrams.

\section{Our approach}

Based on what was discussed in the preceding Section, we write the semiclassical
expression for transport moments as \be\label{semi} M_m(N_1,N_2)=\sum_{m_1,m_2}
\sum_{E,V}\mathcal{D}(m_1,m_2,E,V)\frac{(-1)^V}{N^{E-V}}[N_1]_{m_1}[N_2]_{m_2},\ee where
$\mathcal{D}(m_1,m_2,E,V)$ is the number of labeled diagrams with $E$ edges and $V$
vertices, having $m_1$ ($m_2$) distinct channels on the incoming (outgoing) lead.

In this Section we summarize the approach to the calculation of (\ref{semi}) that was
developed in \cite{epl}, following a rather different presentation. We start by
addressing correlated periodic orbits. Then we discuss how they are turned into
scattering diagrams. Next, we introduce our problem regarding factorizations of
permutations. We then consider the combinatorics of channels and finally present the end
result.

\subsection{Correlated Periodic Orbits}

In \cite{haakepre}, pairs of correlated periodic orbits were associated with
factorizations of permutations. We generalize it to take into account a situation when a
single periodic orbit $\alpha$ is correlated with a set of periodic orbits
$\beta_1,\beta_2,$ etc. In-between encounters, $\alpha$ and $\beta$ are
indistinguishable, and we can label the encounter stretches in such a way that the end of
stretch $j$ is followed by the beginning of stretch $j+1$. This produces the permutation
$c_E$, where $E$ is the number of stretches, acting on the `exit-to-entrance' space (it
goes from the exit of an encounter to the entrance of another one).

\begin{figure}[t]
\includegraphics[scale=0.62,clip]{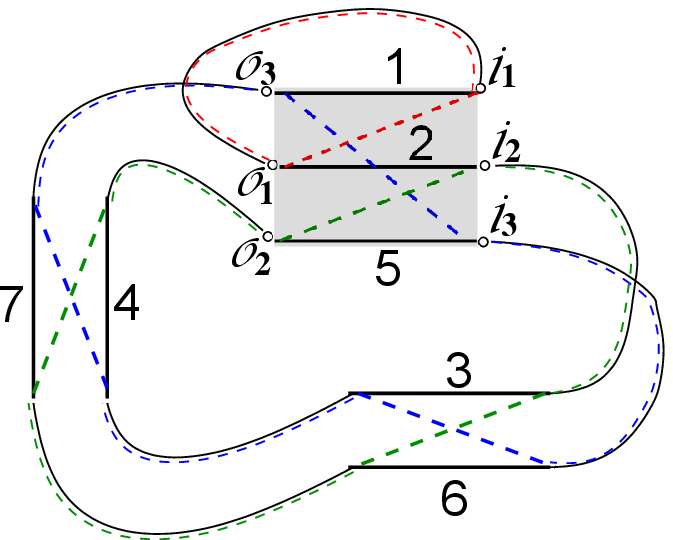}\hspace{0.4cm}
\includegraphics[scale=0.62,clip]{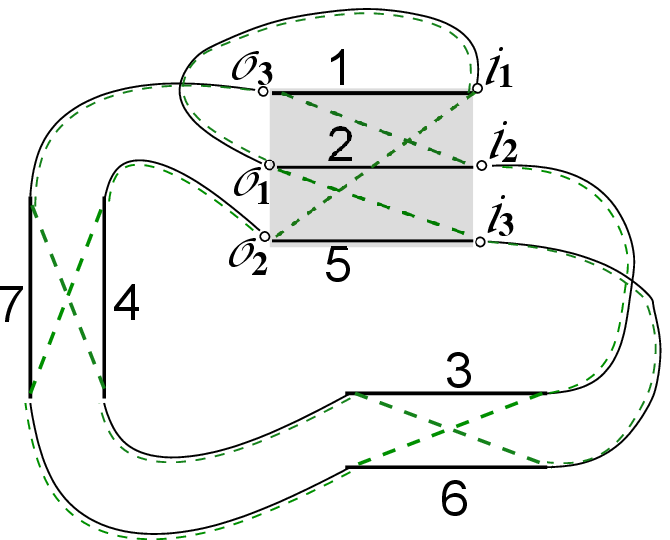}\hspace{0.4cm}
\includegraphics[scale=0.62,clip]{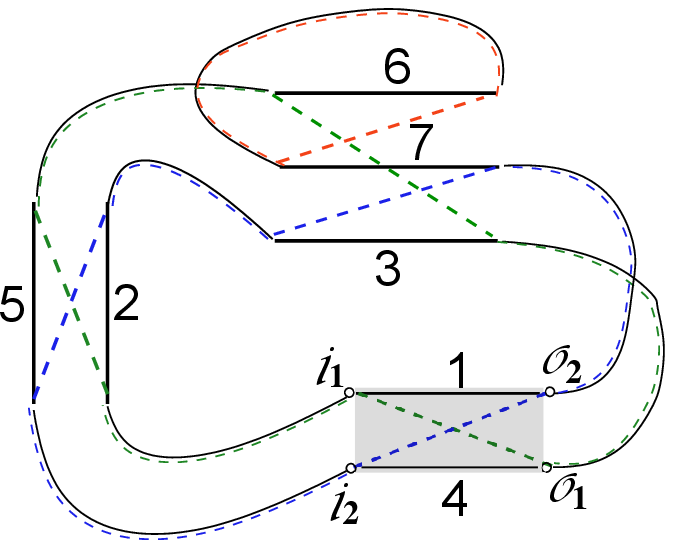}
\caption{(color online) Schematic representation of correlated periodic orbits and how
they become scattering trajectories. Orbit $\alpha$ is depicted with a solid line and
orbits $\beta$ with dashed lines. Shaded regions represent the encounter which is cut
open. a) is associated to the factorization equation
$(1234567)=(1)(264)(375)\cdot(125)(36)(47)$. Notice that the first cycle of $P$ is
increasing. b) is associated to $(1234567)=(1537264)\cdot(152)(36)(47)$; we do not use
these orbits to produce scattering diagrams because the first cycle of $P$ is not
increasing. c) is associated to $(1234567)=(153)(274)(6)\cdot(14)(25)(673)$; we also do
not use these orbits because, when the first encounter is opened, one periodic orbit
remains periodic.} \label{fig1}
\end{figure}

The orbits behave differently inside the encounters (the `entrance-to-exit' space). At
any encounter, the action of $\alpha$ corresponds to the identity permutation: it takes
the entrance of a stretch to the exit of the same stretch. On the other hand, $\beta$
acts by implementing a non-trivial permutation, which for convenience we call $P^{-1}$
(this differs from the notation in \cite{epl}). In the example shown in Figure \ref{fig1}
we have $P^{-1}=(152)(36)(47)$. The encounters are in bijection with the cycles of $P$.
Notice that $P$ does not have fixed points.

The product $c_E P^{-1}\equiv Q$, acts on `exit-to-exit' space, leading from the exit of
an encounter to the exit of another one. Each cycle of $Q$ corresponds to one of the
periodic orbits in $\beta$. Since $c_E$ is fixed, the total number of correlated pairs
equals the number of solutions to the factorization equation $c_E=QP$. As we have seen,
this is counted by connection coefficients.

A few examples are shown in Figure 1. In Fig.1a there are three different $\beta$'s, and
we have $Q=(1)(264)(375)$. In Fig.1b there is only one $\beta$ because the exchange of
partners inside the $3$-encounter is different from Fig.1a. In this case $Q=(1537264)$.
Finally, in Fig.1c we have simply relabeled the stretches of Fig.1a so that the first
stretch belongs to a $2$-encounter.

\subsection{Turning Periodic Orbits into Scattering ones}

Given $\alpha$ and $\beta$, we `cut open' an $m$-encounter (as a matter of convention, we
always open the encounter that contains the stretch labeled `$1$') to produce $2m$
endpoints, $m$ of them corresponding to the `beginning' of trajectories (leaving the
encounter) and the other $m$ to `ending' of trajectories (arriving at the encounter). We
interpret them as incoming and outgoing channels, respectively. Then, the first stretch
becomes $i_1$, and we use $\alpha$ to label all channels in sequence: the piece of
$\alpha$ that starts in $i_j$ (and necessarily ends in $o_j$) becomes $\gamma_j$, while
the piece of $\beta$ that starts in $i_j$ becomes $\sigma_j$. This produces something we
have called a {\em pre-diagram} in \cite{epl}. It resembles a diagram, but there are some
points about which we must be careful.

The first is that if one of the $\beta$ orbits does not participate in the encounter, it
remains a periodic orbit instead of becoming a scattering trajectory. We must therefore
demand that this does not happen. Second, in the resulting situation the $\sigma$
trajectories do not necessarily connect incoming to outgoing channels according to the
cyclic permutation $c_m$. We must enforce some coincidences among the channels so that
the permutation implemented by the $\sigma$'s, call it $\pi$, becomes effectively equal
to $c_m$. Finally, some different $\alpha,\beta$ can lead to the same situation when they
are opened. However, when this happens the structure of the first encounter is
necessarily different. In order to avoid this over-counting, we impose that the
permutation experienced by the $\sigma$'s inside the first encounter must be decreasing.

In Figure 1 the lightly shaded regions indicate the encounters that are opened in order
to produce scattering situations. The pre-diagram which arises from Fig.1a is shown in
Fig.2a. As it is, the $\sigma$ trajectories implement the identity permutation in the
channel labels. In Fig.2b and Fig.2c we show two possible diagrams that can be obtained
from this pre-diagram. Note that trajectories $\sigma_2$ and $\sigma_3$ can be assigned
in two different ways in Fig.2c. Fig.1b does not lead to a pre-diagram, because the
permutation inside the first encounter is not decreasing. This avoids over-counting,
since it would lead to the same pre-diagram as Fig.1a. Fig.1c also does not lead to a
pre-diagram, because it has a periodic orbit which does not participate in the first
encounter.

\begin{figure}[t]
\includegraphics[scale=0.62,clip]{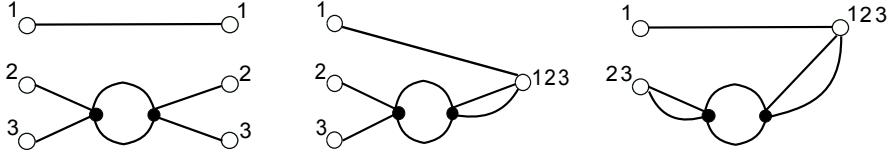}
\caption{a) The pre-diagram obtained when we open the first encounter of the correlated
periodic orbits shown in Fig.1a. This is not a true diagram, because the permutation
implemented by $\sigma$ on the channel labels is $(1)(2)(3)$ instead of $(123)$. A true
diagram can be produced when there are coincidences among the channels, as in b) and c)
for example. Notice that in c) the identity of trajectories $\sigma_2$ and $\sigma_3$ can
be interchanged.} \label{fig2}
\end{figure}

Taking into account the above discussion, we write $\mathcal{D}(m_1,m_2,E,V)$ as \be
\mathcal{D}(m_1,m_2,E,V)=\sum_{\pi\in\S_m}\Xi(m,E,V;\pi)f(\pi,m_1,m_2).\ee Here
$\Xi(m,E,V;\pi)$ is the number of pre-diagrams with $m$ channels on each side, having $E$
edges, $V$ vertices and implementing permutation $\pi$. On the other hand,
$f(\pi,m_1,m_2)$ is the number of ways to convert such a pre-diagram into a labeled
diagram with $m_1$ and $m_2$ distinct channels.

\subsection{A New Factorization Problem}

Since all elements of $\beta$ must take part in the first encounter, all cycles of
permutation $Q$ must have at least one element in common with the first cycle of
permutation $P$. To the knowledge of this author, the problem of factorizing the cycle
under this condition on the factors has not been considered before.

We must also determine what is $\pi$ for a given factorization $c_E=QP$. Let $\{P\}$
denote the set of integers which are not fixed points of the permutation $P$. Let $P_1$
denote the first cycle of $P$, the one that contains the element `$1$'. Given a set $s$,
let $\left.P\right|_s$ denote the {\em restriction} of $P$ to $s$, defined in Section 2.

Suppose we have correlated orbits described by the equation $c_E =QP$. The set of
elements involved in the first encounter is $\{P_1\}$, assumed to have $m$ elements. In
the example of Fig.\ref{fig1}a this is $\{1,2,5\}$. The $\gamma$ trajectories start and
end at this encounter and, by construction, visit these elements in increasing order,
i.e. they implement a permutation which is simply $\left.c_E\right|_{\{P_1\}}$. This is
$(125)$ in Fig.\ref{fig1}a.

We must determine what is $\pi$, the permutation induced by $\sigma$ on those labels.
First, we restrict $Q$ to the appropriate space, $\left.Q\right|_{\{P_1\}}$. In
Fig.\ref{fig1}a this is $(1)(2)(5)$. This acts on exit-to-exit space, i.e. it takes
incoming channels to incoming channels. We multiply it by $P_1$ in order to reverse the
permutation effected inside the first encounter. The result,
$\left.Q\right|_{\{P_1\}}P_1$, takes incoming channels to outgoing channels. In
Fig.\ref{fig1}a this is also $(125)$, just like for the $\gamma$'s.

At this point, we have the permutations implemented by both $\gamma$ and $\sigma$ on the
channel labels. The first is $\left.c_E\right|_{\{P_1\}}$ and the second is
$\left.Q\right|_{\{P_1\}}P_1$. The first should be the identity, so we multiply both
quantities by $\left.c_E^{-1}\right|_{\{P_1\}}$ to get $\widehat{\pi}$: \be\label{xi}
\widehat{\pi}=\left.Q\right|_{\{P_1\}} P_1\left.c_E^{-1}\right|_{\{P_1\}}.\ee We can
choose the permutation experienced by the $\sigma$'s inside the first encounter be
decreasing, so that $P_1$ is increasing, i.e. $P_1=\left.c_E\right|_{\{P_1\}}$. This
leads to \be \widehat{\pi}=\left.Q\right|_{\{P_1\}}.\ee In particular, the number of
cycles of $\widehat{\pi}$ equals the number of individual periodic orbits in the set
$\beta$. In Fig.1a, $\widehat{\pi}=(1)(2)(5)$.

Finally, notice that $\widehat{\pi}$ is a permutation of $m$ symbols, but these are not
the elements of the set $\{1,\ldots,m\}$ (as we have seen in the example). We must now
represent $\widehat{\pi}$ as a permutation acting on this set. This is achieved by making
every element of $\widehat{\pi}$ as small as possible while maintaining relative order.
That is, the element $1$ remains the same while the next larger element becomes $2$, etc.
Denote this operation by $\langle \bullet\rangle$. For example, $\langle
(13)(475)\rangle=(12)(354)$. The permutation $\pi$ is then $\langle
\widehat{\pi}\rangle$.

Let $\Xi(m,E,V;\pi)$ be the number of solutions in the permutation group $S_E$ to the
factorization equation $(12\cdots E)=QP$ which satisfy the following conditions: i) $P$
has $V+1$ cycles and no fixed points; ii) its first cycle, $P_1$, is increasing and of
size $m$; iii) all cycles of $Q$ have at least one element in common with $P_1$; iv)
$\left\langle Q_{\{P_1\}}\right\rangle=\pi$.

Obviously, the problem is only defined for $m\leq E$. If $m=E$, then necessarily $V=0$
and in that case $\Xi(m,m,0,\pi)=1$ since the only solution is $c_m=1\cdot c_m$. Since
$P$ has no fixed points, the largest possible value for $V$ is the integer part of
$(E-m)/2$.

It seems very natural that $\Xi(m,E,V;\pi)$ should depend on $\pi$ only via its cycle
type, and we have strong numerical evidence in favour of that. This is what motivates our
Definition 1, which appears in Section 2. Assuming this is indeed true, then clearly \be
\label{pitoo}\Xi(m,E,V;\omega)=|\C_\omega|\Xi(m,E,V;\pi),\ee if we denote by $\omega$ the
cycle type of $\pi$.

\subsection{Combinatorics of channels}

A pre-diagram only becomes an actual diagram if certain coincidences exist among the
channels so that $\pi$ is effectively equal to $c_m$, as required by $(\ref{bigsum})$. We
thus consider the following problem: Given a pre-diagram with $m$ channels on both sides,
in how many ways can we turn it into an acceptable labeled diagram that has $m_1$
distinct channels on the incoming lead and $m_2$ in the outgoing one? The answer to this
question is the function $f(\pi,m_1,m_2)$ which we have already introduced. The
calculation of this function was left as an open problem in \cite{epl}; here we present
its solution.

Take the simplest case first, when $\pi=c_m$ is already the correct permutation and the
pre-diagram is already a true diagram. The idea is to consider all possible permutations
that can be implemented on the $\sigma$'s, and for each such permutation to determine
under which conditions it can be accepted. The answer is easy: if the permutation
implemented has $k$ cycles, then there must be at most $k$ distinct channels on each
side. The number of permutations of $m$ symbols with $k$ cycles is equal to $s(m,k)$. The
number of ways to distribute $k$ distinct numbers among $m_1$ possibilities is
$S(k,m_1)$. We conclude that \be\label{fcm} f(c_m,m_1,m_2)=
\sum_{k=1}^{m}s(m,k)S(k,m_1)S(k,m_2).\ee

For example, take $m=2$ and $\pi=(12)$. With a given labeling, we can have all channels
different, one coincidence of the left, one one the right or coincidence on both sides.
On the other hand, the only possibility in order to be able to interchange the $\sigma$'s
is if there are coincidences on both sides.

Now let us look at the more general situation, when $\pi\neq c_m$. Now, instead of
looking at permutations of the $\sigma$'s, we must consider separately two permutations,
call them $\lambda$ and $\rho$, acting on the left and right channel labels,
respectively. We count all possible such pairs, under the condition that they make $\pi$
effectively equal to $c_m$. This is done as follows. Given any label on the left, say
$i$, the permutation $\lambda$ takes it into another label, $\lambda(i)$. Then
permutation $\pi$ acts, taking it from the left lead to the right lead. At the right
lead, permutation $\rho$ acts and must take the result into $i+1$. In short, we must have
\be\label{decomp} \lambda\pi\rho=c_m.\ee

As an example, consider $\pi=(132)$. One solution is $\lambda=(13)$, $\rho=(13)$. If
$\pi=c_m$, equation (\ref{decomp}) requires $\rho=\lambda^{-1}$.

The number of solutions to $c_m=\lambda\pi\rho$ is the same as the number of solutions to
$c_m\pi^{-1}=\lambda\widetilde{\rho}$, where $\widetilde{\rho}=\pi\rho\pi^{-1}$. Notice
that $\widetilde{\rho}$ and $\rho$ have the same cycle type. If we require that $\lambda$
has $\ell$ cycles and $\rho$ has $r$ cycles, then there are $C^{\alpha}_{\ell,r}$
solutions, where $\alpha$ is the cycle type of $c_m\pi^{-1}$.

By considering all solutions to (\ref{decomp}) we are in fact considering all possible
labelings of the $\sigma$'s. To accept a certain choice of labeling, the left labels that
belong to the same cycle of $\lambda$ must belong to the same channel. Similarly, the
right labels that belong to the same cycle of $\rho$ must also belong to the same
channel. If $\lambda$ has $\ell$ cycles and $\rho$ has $r$ cycles, there are
$S(\ell,m_1)S(r,m_2)$ ways to arrange this in the leads. Therefore, in view of the above
paragraph, we have \be
f(\pi,m_1,m_2)=\sum_{\ell,r=1}^mC^{\alpha}_{\ell,r}S(\ell,m_1)S(r,m_2).\ee

Coming back to the semiclassical expression for moments, equation (\ref{semi}), we see
that there appears the sum \be F(\alpha,N_1,N_2)=\sum_{m_1,m_2=1}^m
f(\pi,m_1,m_2)[N_1]_{m_1}[N_2]_{m_2}.\ee According to basic properties of Stirling
numbers, reviewed in Section 2, this is given by \be\label{FN}
F(\alpha,N_1,N_2)=\sum_{\ell,r=1}^mC^{\alpha}_{\ell,r}N_1^\ell N_2^r.\ee

\subsection{Final Result}

In terms of previously defined quantities, our semiclassical approach yields the
following result for transport moments:\be\label{final1}\fl M_m=\sum_{\pi\in S_m}
F(\alpha,N_1,N_2)\sum_{d=0}^\infty\frac{1}{N^{m+d}}\sum_{V=0}^d
(-1)^V\Xi(m,V+m+d,V;\pi),\ee where $\alpha$ is the cycle type of $c_m\pi^{-1}$ and we
have used $d=E-V+m$ instead of $E$. As we have seen in Section 2, there are
$C^{(m)}_{\omega,\alpha}$ many permutations $\pi$ with cycle type $\omega$ such that
$c_m\pi^{-1}$ has cycle type $\alpha$. Therefore, assuming (\ref{pitoo}) we can also
write \be\label{Mome2}\fl M_m=\sum_{\alpha,\omega\vdash m}
\frac{C^{(m)}_{\omega,\alpha}}{|\C_\omega|}
F(\alpha,N_1,N_2)\sum_{d=0}^\infty\frac{1}{N^{m+d}}\sum_{V=0}^d
(-1)^V\Xi(m,V+m+d,V;\omega).\ee Let us introduce \be
C^{(m)}_{\omega,\ell,r}=\sum_{\alpha\vdash
m}C^{(m)}_{\omega,\alpha}C^{\alpha}_{\ell,r}\ee which is the number of solutions to the
factorization $c_m=\rho\pi\lambda$ where $\lambda$ has $\ell$ cycles, $\rho$ has $r$
cycles and $\pi$ has cycle type $\omega$. Using this quantity and the relation
(\ref{FN}), we get \be\label{final2} \fl
M_m=\sum_{\ell,r=1}^m\sum_{d=0}^\infty\frac{N_1^\ell N_2^r}{N^{m+d}}\sum_{\omega\vdash m}
\frac{C^{(m)}_{\omega,\ell,r}}{|\C_\omega|} \sum_{V=0}^d(-1)^V\Xi(m,V+m+d,V;\omega).\ee

\section{Comparisons with other results}

Next, we compare the above results with RMT and with the semiclassical approach of
\cite{also} in order to derive the conjecture about the function $\Xi(m,V+m+d,V;\omega)$
that was stated in Section 2.

\subsection{Other semiclassics}

The semiclassical calculation of transport moments was also considered in \cite{also},
along different lines. In particular, a function $p_d(\pi)$ is required which is the
number of primitive factorizations of depth $d$ that exist for the permutation $\pi$.
This function is interesting on its own and is further discussed in \cite{merola,novak}.

For systems with broken time-reversal symmetry, the result obtained in \cite{also} for
the unitary symmetry class reads \be\fl M_m(N_1,N_2)= \sum_{\lambda,\rho \in
\S_m}N_1^{c(\lambda)}N_2^{c(\rho)}\sum_{d=0}^\infty\frac{1}{N^{m+d}} (-1)^dp_d(\lambda
c_m\rho),\ee where $c(\bullet)$ denotes the number of cycles of a permutation. This can
be shown to be exactly equivalent to RMT. The function $p_d(\pi)$ depends only on the
cycle type of $\pi$, which we denote $\omega$.

Using the quantity $C^{(m)}_{\omega,\ell,r}$ already defined, this can be written as \be
M_m=\sum_{\ell,r=1}^m\sum_{d=0}^\infty\frac{N_1^\ell N_2^r}{N^{m+d}}\sum_{\omega\vdash m}
C^{(m)}_{\omega,\ell,r}(-1)^d p_d(\omega). \ee Since this expression must agree with our
result (\ref{final2}) for all values of $N_1,N_2$, we arrive at our Conjecture, stated in
Section 2.

At the heart of the approach developed in \cite{also} is a technique for taking as much
advantage as possible from cancelations between diagrams. Our approach, embodied in the
function $\Xi(m,E,V,\omega)$, does not have this characteristic. The conjecture we
present here therefore encodes diagram cancelations. Perhaps taking into account other
degrees of cancelations might lead to other factorization problems.

\subsection{RMT with $N_2=1$}

Let us consider the case when the right lead is in the extreme quantum regime $N_2=1$.
Then \be F(\alpha,N_1,1)=\sum_{\ell,r=1}^mC^\alpha_{\ell,r}N_1^\ell N_2^{r}
=\sum_{\ell=1}^ms(m,\ell)N_1^\ell=[N_1]^m,\ee where we used (\ref{ris}). Using that
$\sum_\alpha C^{(m)}_{\omega,\alpha}=|\C_\omega|$, equation (\ref{Mome2}) gives
\be\label{Mn2}\fl
M_m(N_1,1)=[N_1]^m\sum_{d=0}^\infty\frac{1}{(N_1+1)^{m+d}}\sum_{\omega\vdash
m}\sum_{V=0}^{d}(-1)^V \Xi(m,V+m+d,V;\omega).\ee Using the consequence (\ref{conseq}) we
derived from our Conjecture, this becomes \be\label{toRMT}\fl
M_m(N_1,1)=[N_1]^m\sum_{d=0}^\infty\frac{1}{(N_1+1)^{m+d}}(-1)^dS(m+d-1,m-1)
=\frac{[N_1]^m}{[N_1+1]^m}.\ee

The RMT prediction for transport moments in the absence of time-reversal symmetry is
\cite{prb78mn2008}\be M_m(N_1,N_2)=\sum_{p=0}^{m-1}\frac{(-1)^p}{m!}{m-1 \choose
p}\frac{[N_1-p]^m[N_2-p]^m}{[N_1+N_2-p]^m}.\ee When $N_2=1$, we may use
$[1-p]^m=m!\delta_{p,0}$ and this reduces exactly to (\ref{toRMT}).

\section{Conclusions}

We have derived a semiclassical expression for transport moments, valid for arbitrary
numbers of channels and in agreement with random matrix theory for broken time-reversal
symemtry (as far as it can be checked). Unfortunately, the combinatorial problem of
determining $\Xi(m,E,V;\omega)$ is still open, so we cannot show exact agreement with
RMT. However, the merit of the semiclassical approach is not in its computational
efficiency, but rather as a way to identify the dynamical origins of universality.

A deep connection has been revealed between quantum chaotic scattering and the problem of
factorizing permutations (see also \cite{preprint}). In fact, two different factorization
problems have appeared in this area, the one discussed here and the one discussed in
\cite{also} (factorizations of permutations had already appeared \cite{haakepre} in the
semiclassical approach to closed chaotic systems). These two problems are quite different
in nature, but there probably is a relationship between them, suggested by the conjecture
we have put forth in the present work. This remains to be further investigated.

RMT statistics is expected to hold when the average dwell time in the cavity is much
larger than the system's Ehrenfest time. This was assumed in the present work. When these
two time scales are comparable, the semiclassical approach is more complicated, but some
results have been obtained \cite{TE1,TE2,TE3}. Whether these can be extended to a
complete calculation of all moments, and what kind of combinatorics is required, is still
an open problem. Other interesting open problems include the treatment of tunnel barriers
(see \cite{jack1,jack2}) or Andreev reflection (see \cite{andreev1,andreev2}). Finally,
diffraction effects should be important when the number of channels is small
\cite{rotter}, and this is yet to be considered.

\ack I would like to thank Jean-Gabriel Luque for important correspondence in the early
stages of this work. Financial support was provided by FAPESP and CNPq.

\end{document}